\documentclass{article}

\usepackage{arxiv}

\usepackage[utf8]{inputenc} 
\usepackage[T1]{fontenc}    
\usepackage{hyperref}       
\usepackage{url}            
\usepackage{booktabs}       
\usepackage{amsfonts}       
\usepackage{nicefrac}       
\usepackage{microtype}      
\usepackage{lipsum}		
\usepackage{graphicx}
\usepackage{natbib}
\usepackage{doi}

\title{Review Of Integrated Photonic Elastic WDM Switches For Data Centers}

\author{ {Akhilesh~S~P~Khope}\\
	Microsoft Corporation\\
	One Microsoft Way\\
	Redmond, WA, 98052\\
	\texttt{akhkhope@gmail.com} \\
	\And
	{Anirban Samanta, Xian~Xiao, Ben~Yoo} \\
	Department of Electrical And Computer Engineering\\
	One Shields Avenue, Kemper Hall\\
	University of California\\
	Davis, CA 95616\\
	\texttt{} \\
	\And
	{John~E~Bowers} \\
	Department of Electrical And Computer Engineering\\
	University of California\\
	Santa Barbara, CA, 93106\\
	\texttt{} \\
}

\renewcommand{\shorttitle}{\textit{arXiv} Template}

\hypersetup{
pdftitle={A template for the arxiv style},
pdfsubject={q-bio.NC, q-bio.QM},
pdfauthor={David S.~Hippocampus, Elias D.~Striatum},
pdfkeywords={First keyword, Second keyword, More},
}

\begin{document}
\maketitle

\begin{abstract}
In this review paper, we present an elaborate discussion on wavelength selective switches and their demonstrations. We also review packaging and electronic photonic integration of switches; a topic neglected in other review papers. We also cover wavelength locking which is paramount in switching networks with many tunable filters. 
\end{abstract}


\section{Introduction}
Switch is a device used to turn off and on a connection between two ports. If N terminal devices like telephones or computer servers are connected to each other, switch can reduce the number of connections from $N(N-1)$ to $2N$ and also intelligently provision bandwidth according to demand. One of the first demonstrations of a crossbar switch was in 1915 \cite{craft1915multiple}, the switch was an electromechanical switch used for switching telephone calls. Widespread use of crossbar switches is reported in Sweden in 1926 and later in 1938 in AT\&T Bell Labs. 

Larger port count switches are hard to manufacture and thus Charles Clos created an ingenious way of creating a bigger network from smaller switches in 1953 \cite{clos1953study}. Many datacenter architectures follow a modified version of this architecture called fat tree topology. Switching networks as well as switches which form them can be either blocking, strictly non blocking or reconfigurable non blocking. For a detailed discussion on this, reader is advised to read \cite{benevs1965mathematical}. In this paper, many of the integrated photonic switches reported are rearrangeably non-blocking as their configuration changes according to an arbitrator on a global clock. With the miniaturization of devices in Silicon Photonics, advances in lasers and Semiconductor Optical Amplifiers (SOAs) in III-V platforms and tight electronic and photonic integration through fabless photonic foundries it is now possible to fabricate and integrate larger port count switching networks on a single chip.

The communication patterns in the today's high performance computing and datacenter systems are spatially and temporally non-uniform, which means the bursty bandwidth requirement in some data links could cause heavy congestion. However, today's electronic switches based interconnections have a fixed topology, which is incapable of dynamically adapting the bandwidth to the workloads. In recent years, several optical bandwidth-reconfigurable switching fabrics have been demonstrated by using wavelength-and-space selective optical switches. By utilizing wavelength division multiplexing (WDM) technologies, such switches can achieve bandwidth steering by reconfiguring the number of wavelength channels in selected pairs of nodes while maintaining the benefit of optical switching, including higher speed and lower power consumption.

Architecture of most of the switches reported in this review consists of a wavelength division multiplexed (WDM) signal at the input of an integrated photonic switch and a receiver with a demux. 
Integrated photonic switches can be categorized into broadband/fiber switches and wavelength selective switches (WSS). Wavelength selective switches can be further categorized into single wavelength selective and elastic/multiwavelength selective switches. In broadband switches, the entire WDM signal from one port is switched to another port. In single wavelength selective switches connection between ports has only one wavelength connection between ports, and if one changes the input wavelength at input port the signal is received at a different output port. In elastic/multiwavelength selective switch, different ports can be connected to each other using one or more than one wavelength from the WDM signal. 


In the following two paragraphs, we report some data center architectures with WSS from literature. Such switches can be integrated into datacenter networks and high performance computing clusters with some modification. Fig. 1(a) shows a data center network with multi degree WSS mesh connecting different server groups reported by Google in \cite{zhao2016dynamic}. The optical switches in the WSS mesh can increase network scalability and result in fewer physical links between elements of the data center network which results in reduced hardware and network cost. Fig. 1(b) shows a datacenter architecture with WSS in a multidimentional interconnection network called Hyper-X is reported in \cite{saleh2016elastic}. Hyper-X requires very dense fiber connections for large radix HPC and datacenter networks and the induction of WSS switches enables reconfigurability and results in simplifying wiring. 

Fig. 1(c) shows a method of designing networks based on AWGs and WSSs is reported in \cite{lin2017construction}. This method is more flexible in terms of construction of different switch sizes. Another method for constructing such networks is reported by the same group in \cite{huang2016new}. Fig. 1(d) shows a switch with broadband photonic switch and WSS for switching between top of the rack switches is reported in \cite{xu2011hybrid}. Fig. 1(e) shows an AWGR based architecture for datacenters (citation). Fig. 1(f) shows an Optical Top Of the Rack (OTOR) architecture that can be integrated into a software defined network (SDN) enabled datacenter architecture with dynamic bandwidth allocation using WSS \cite{xue2020sdn}. Fig. 1(g) shows application of IRIS WSS used as a transponder aggregator for Reconfigurable Optical Add Drop Multiplexers (ROADMs) \cite{testa2018integrated}.

\begin{figure*}[htbp]
\centering
\includegraphics[width=\linewidth]{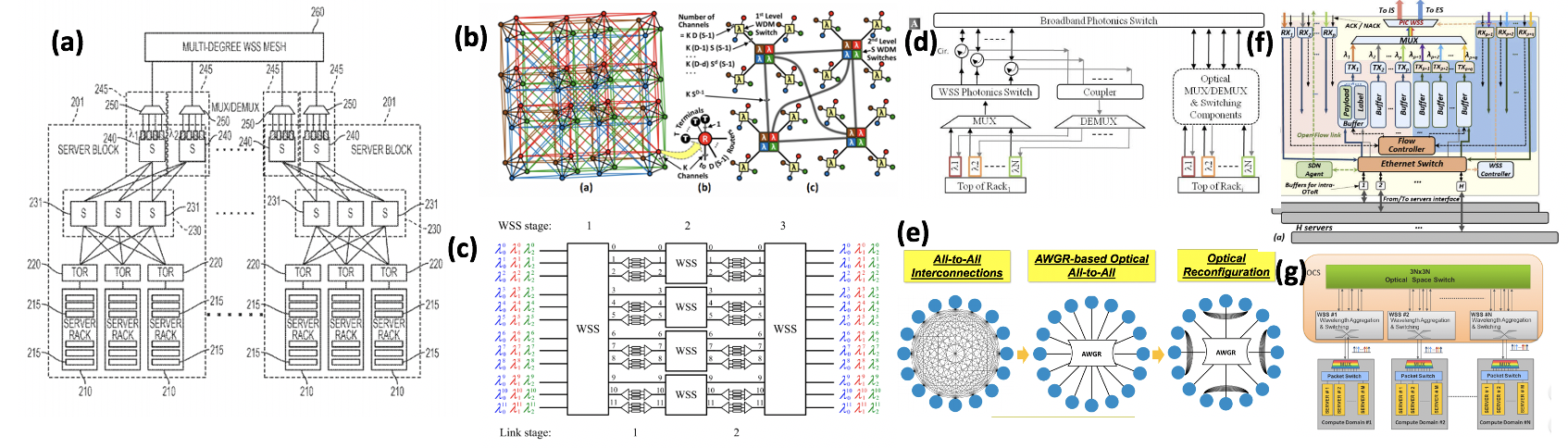}
\caption{(a) check this again , Dynamic datacenter network with Multi-Degree WSS mesh, (b) HyperX topology for WSS (bigger) , (c) Non-blocking wavelength space switches with WSSs \cite{lin2017construction}, (d) switching between top of the rack switches, (e) application of AWGR based switches in datacenters, (f) photonic top of the rack switch (g) application of IRIS WSS ROADMs
}
\end{figure*}

\begin{figure*}[htbp]
\centering
\includegraphics[trim=0 0 5
0,clip,width=\linewidth]{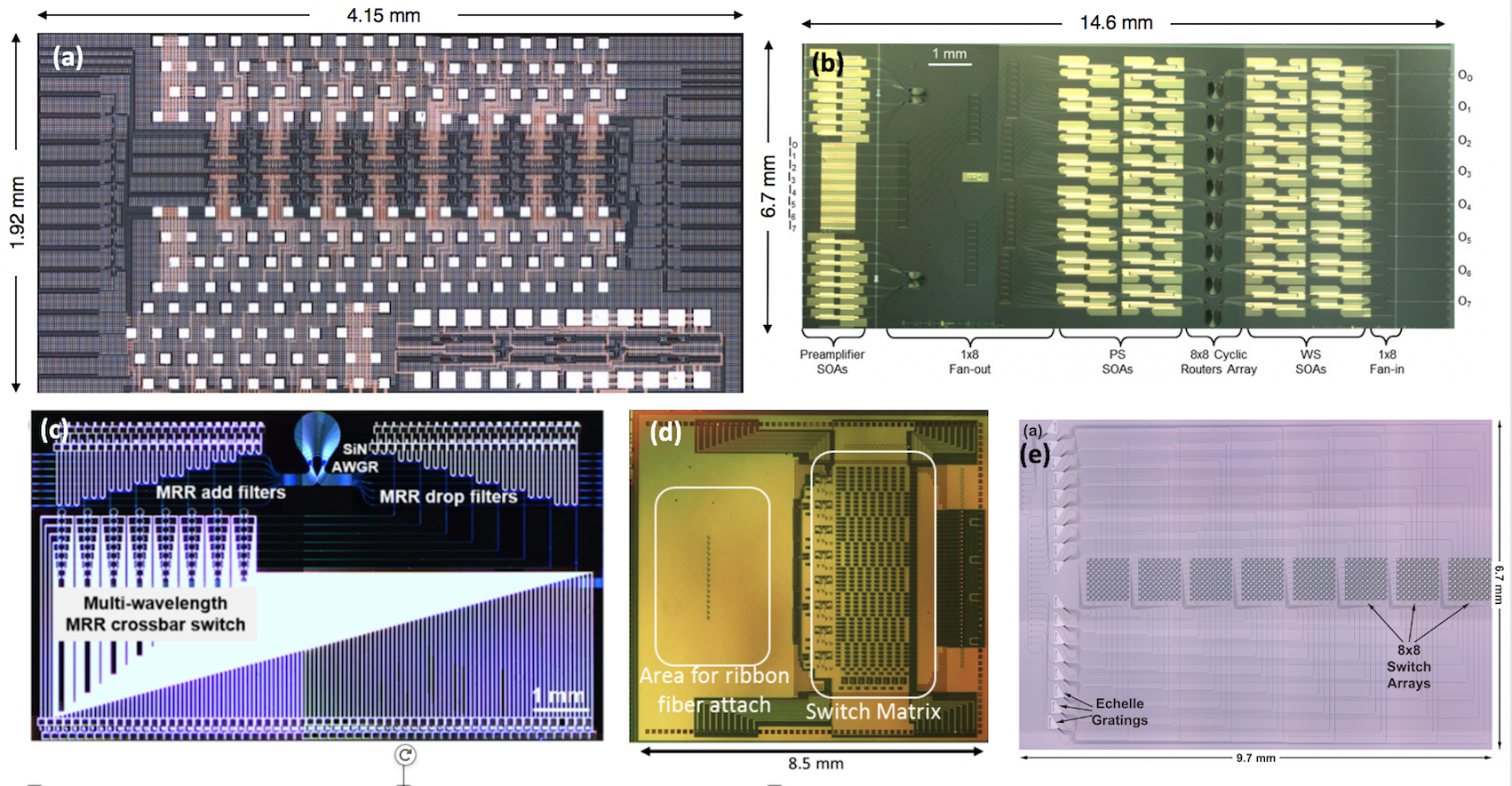}
\caption{(a) 8x4 switch reported in \cite{khope2019multi}, (b) 8x8 SOA based switch reported in \cite{stabile2013monolithically}, (c) Flex lions based switch reported in \cite{xiao2019flex}, (d) Reconfigurable silicon photonic ROADM is reported in \cite{testa2016design,testa2018integrated}, (e) $8 \times 8 \times 8$ SiP MEMS switch is reported in \cite{seok2019silicon}}
\end{figure*}


There have been numerous reviews on photonic switching and datacenter technologies \cite{cheng2018recent,lee2017silicon,shen2019silicon,cheng2018photonic}, none of them have covered integrated photonic elastic WDM switches.



The main difference between our paper and the other recently published review papers is:
\begin{enumerate}
    \item we do an in depth review of integrated photonic elastic WDM switches as compared to other papers.
    \item we do an in depth review of packaging and integration of the switches with electronics. 
    \item we also cover wavelength locking and stabilization.
\end{enumerate}


In Section II of this review, we review integrated photonic architectures. In Section III, we compare different switching devices. In Section IV, we review demonstrations of wavelength selective switches. 
In Section VI, we discuss packaging technologies for both broadband and WSS. 
In Section VII, we discuss wavelength locking and stabilization. 

\section{Architectures}
\label{sec:headings}
Smaller switches can be arranged in different architectures according to application. WSS can be arranged in the following architectures (Maybe diagram):
\begin{itemize}
    \item Layered architecture
    \item Crossbar 
    \item AWGR based switches
\end{itemize}

All switching architectures have N input ports , an output ports and M wavelengths at each input.

Crossbar switch: In this arrangement there are upto M ring resonators at each intersection.  There are $MN^2$ switching elements. Optical path loss is path dependent. 

Layered switch  : In this arrangement, N demultiplexers routes M wavelengths to M  $N\times N$ single wavelength switches and these are connected to N multiplexers. Single wavelength switches in this arrangement can be AWGRs, Crossbar switches, Benes switches and any other non blocking switching arrangement. This type of switch can be built in a silicon photonic platform or III-V platform using SOAs. There are $MN^2$ switching elements if using crossbar as a switching element. $M + 2N$ switching elements if using cyclic AWGRs, multiplexer  and demultiplexer .$MNlogN$ switching elements is a Benes network is used.Authors in \cite{wagner1996monet} reported MONET, a layered switch for metro and long haul networks.

AWGR based architecture:  
In this arrangement, M tunable lasers are used at each N ports. A cyclic AWGR is used to route wavelengths to different ports based on the input wavelength from tunable lasers. There are $2MN$ active elements (tunable lasers). 

\section{Demonstration of switches}
In multi-wavelength switches, in addition to all to all extra connections can be provisioned between nodes as per demand. This reconfiguration provides superior latency when load is higher as compared to fixed connectivity all to all switches. These switches also require a control plane for arbitration. Multi-wavelength crossbar MRR based switches fabricated in silicon photonics foundries are reported in \cite{khope2017chip,khope2019multi}. The authors switch in \cite{khope2016elastic} and we carry out architectural exploration in  \cite{saleh2016elastic}. These switches have upto two wavelength connectivity per connection and provide latency comparable to that of full wavelength connectivity. Cyclic Arrayed Waveguide Grating Router (AWGR) based switches are reported in Flex Lions \cite{xiaoFlexLions}. These family of switches use two FSRs of the AWGR, to provide multi-wavelength connectivity. Fully flexible switches are reported in MRR based Benes network \cite{goebuchi2008optical} and a layered wavelength selective 8x8x8 MEMS switch reported in \cite{seok2018mems}. A switch and select switch is reported in Polarization-Diversity MRR based switch fabric \cite{yang2020polarization}. A crossbar switch capable of joint unicast and multi-cast functionality  is reported in \cite{su2015wavelength}. In this switch architecture an all to all connectivity is provided by partial drop MRR and full connectivity is provided on demand by a full power drop MRR. A 2x2 2 channel WSS with unlimited free spectral range using contra directional grating couplers is reported in \cite{ikeda2020silicon}. A multicast and multi wavelength selective switch is reported \cite{khope2021scalable,khope2021scalableConf}.
Multi-wavelength selective switch experiments are also reported in \cite{experimentsMSCS}. Push—pull microring-assisted space-and-wavelength selective switch crossbar switch is reported in \cite{huang2020push}. 

A $1 \times 2$ WSS MRR based on nested pairs of subrings is demonstrated in \cite{wu2015compact}. A $1 \times 2$ wavefront control type WSS was reported using silicon waveguides in \cite{Nakamura:18}. A $2 \times 2$ WSS with four wavelength channels, thermo optic MZ switches and MRRs with a wavelength transmittance change of 9.7 dB is reported in \cite{miura2012silicon}. A silicon based $1\times 2$ WSS with foldback AWG is reported in \cite{nakamura2018silicon}. A $4 \times 4 \times 4$ hitless MRR based non blocking switch is reported in \cite{goebuchi2008optical}. A monolithic $1\times 2$ WSS gridless switch in silicon and silicon nitride waveguides is reported in \cite{doerr2011monolithic}. A 200 GHz, 17 channel $1\times 2$ SiP WSS with AWG loopback is reported in \cite{asakura2015200}. This switch reported losses from 21 dB to 26 dB and a crosstalk range of -21 to -2 dB.  A flexible grid $1\times2$ WSS in SiP with 9 MRR is reported in \cite{wang2017design} with a crosstalk lower than -11.5 dB and a maximum tuning bandwidth of 4.09 nm.

A 2x2 microring based WSS is reported in \cite{xu2011hybrid, lira2009broadband}. A scalable 2x2 crossbar silicon photonic switch is reported in \cite{li2015single}. A SiP mode and wavelength switch is reported in \cite{velha2015silicon}. A high speed 2x2 SiP multi-wavelength switch is reported in \cite{lee2009high}, maybe similar paper \cite{lee2008high}. A MRR based cascaded SiP ring crossbar switch is reported in \cite{poon2008cascaded}. An AWGR based WSS is reported in \cite{yu2013scalable}. A slow light based WSS using photonic crystals is reported in \cite{beggs2008ultracompact}. A thermo optic photonic crystal based switch 2x2 is reported in \cite{zhou2017compact}.Wavelength-selective $2 \times 2$ optical switch based on a Ge2Sb2Te 5-assisted microring \cite{zhang2020wavelength}. A $2\times2\times2\lambda$ switch is reported in \cite{huang2020push}. A layered switch with switch and select stages is proposed in \cite{cheng2019scalable}.

InP based switches have a larger footprint for the same switch radix and are typically built using SOAs. These switches can have zero net optical loss.
A 8x8 wavelength and space cross-connect is reported in InGaAsP/InP platform with $1\times8$ broadband selection stages and $8\times8$ gated cyclic routers in \cite{stabile2013monolithically}.
A $1\times8$ optical module of a $8\times8\times8\lambda$ wavelength selective switch and select architecuture with SOAs in InP is reported in \cite{prifti2018performance}.
SiP switches with integrated SOAs can provide zero optical path loss and are reported in \cite{konoike2018soa}. 
A tunable 2x2 WSS in SiO2-Si3N4 is reported in \cite{geuzebroek2005compact}. AWGR in flex lions is on top of silicon.

\section{Packaging}
Packaging of photonic circuits requires consideration of optical, electrical, thermal and mechanical challenges of integrating photonic circuits to off-chip interfaces. In terms of signal transmission the two main domains are the optical fiber-to-waveguide coupling and integration with electronics.

\begin{figure*}[htbp]
\centering
\includegraphics[width=\linewidth]{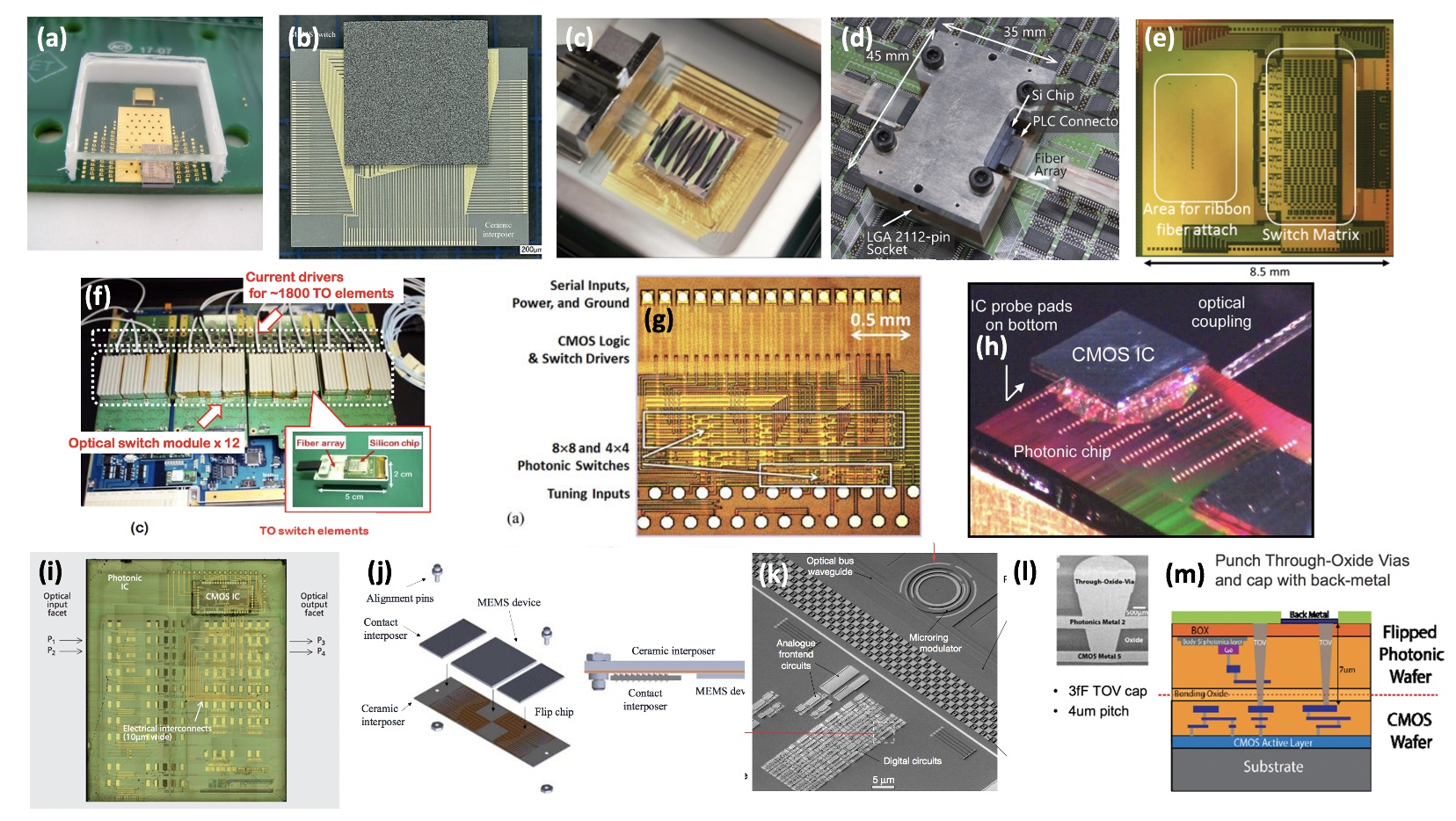}
\caption{(a) Switch die integrated onto a PCB with overhang for edge coupling as reported in \cite{khope2019multi}, (b) MEMS switch flip-chip packaged on a ceramic package reported in \cite{hwang2017flip}, (c) SiPh switch die wirebonded in a CBGA package reported in \cite{dumais2017silicon}, (d) Switch die mounted on a PCB with control electronics reported in \cite{suzuki2019low}, (e) Switch PIC with grating couplers for vertical coupling of light reported in \cite{testa2018integrated}, (f) A transponder system with switch module packaged with fiber array reported in \cite{nakamura2016optical}, (g) Monolithic photonic switch system-on-chip reported in \cite{lee2013monolithic}, (h) CMOS driver IC flip-chip bonded onto photonic chip for 3D vertical integration as reported in \cite{rylyakov2011silicon}, (i) Heterogeneous integration of CMOS die into a cavity of a PIC die followed by further processing to establish metallic contacts as reported in \cite{chen2013electronic}, (j) A pluggable SiPh switch concept reported in \cite{hwang2018pluggable}, (k) Monolithic integration of active photonic devices reported in \cite{atabaki2018integrating}, (l) Through-oxide via (TOV) technology for vertical integration of electronic interconnect and (m) Backside processing of flipped photonic die to form TOVs reported in \cite{settaluri2015} and \cite{stojanovic2015high}}
\end{figure*}

\subsection{Fiber-to-waveguide Coupling}
\subsubsection{Grating Coupler}
A Bragg diffraction grating structure of periodic or chirped design is utilized to vertically couple light to and from an off-chip fiber, making this technique wavelength dependent, resulting in narrow bandwidth, wavelength and polarization sensitivity. They tend to be compact, have high alignment tolerance, simpler wafer-scale testing and high volume manufacturing. Fabricated grating couplers with more than 70\% measured coupling efficiency have been demonstrated \cite{Marchetti2017}, \cite{Kang:20}.

\subsubsection{Edge Coupler and Evanescent Coupler}
An edge coupler consists of an inverse taper section where the waveguide width is gradually reduced to a fine tip. As the guided wave propagates along the taper, it becomes less confined and it's effective cross-section increases. The expanded wave can then be coupled into a lensed fiber. Edge coupling can be more efficient than grating couplers, with up to 90\% efficiency \cite{takei2013}, however they require highly polished facets to achieve high efficiency necessitating complex fabrication processes. Usually long tapers are required for adiabatic functioning, and they also suffer from mode mismatch challenges. Evanescent coupler is an extension of the edge coupling concept. Instead of coupling light into a fiber at the edge, the expanded mode is coupled into another tapered waveguide fabricated in another PIC or interposer with optical fibers pre-attached \cite{Dangel:15}. It offers the advantages of edge coupling along with relaxed alignment tolerances characteristic of grating couplers.

\subsection{Electronic Packaging}
Photonic circuits need to be driven by electronics and the high speed nature of the signals require careful design of the electrical interconnect in order to control the impedance values and electrical crosstalk. Electrical pads on photonic chips are connected to electrical pads on driver asics, and these connections are finally connected to the PCB. The pad pitch of both photonic and electronic chips is much smaller than pad pitch of PCB. The mismatch in pitch between PCB and chip can be circumvented by using an intermediary interposer or package with an organic or ceramic ball grid array (BGA). The interconnection is usually done with wire bonds or through vertical integration methods.

\subsubsection{Wire bonding}
If the number of electrical connections is low (typically less than 400), wire bonding offers a simple solution where bare metal wires are used to connect PIC contact pads on the top surface to a leadframe on an organic or ceramic BGA package \cite{Dumais:18}. These packages typically support $150 \, \mu m$ or more in contact pitch with the backside ball grid array pitch being 0.5 mm or more. Wire bonds tend to have higher impedance values limiting electrical performance. 

\subsubsection{Vertical integration}
When a large number of contacts, dense electrical I/O or higher electrical performance is necessary, vertical integration becomes critical. Vertical integration with flip-chip bonding or by using through silicon vias (TSVs) and through oxide vias (TOVs) can reduce the length of electrical interconnect to the order of the IC thickness (100s of $\mu m$) compared to many millimeters in wire-bonded solutions, reducing the parasitic impedance \cite{settaluri2015} allowing for improved performance. In a flip-chip solution, the IC die is flipped with C4 solder bumps or copper pillars used to attach to a traditional organic or ceramic package. However when access to the active side of the die is necessary, for example when using grating couplers, flip-chip bonding may not be possible. In such cases vertical integration is made possible by using TSVs/TOVs where a via is formed through the bulk silicon or oxide from the inner contact pad to a backside metal pad for C4 or copper pillar bonding \cite{bogaerts2018}.

An 8$\times$8 SiPh Flex-LIONS chip with with 176 electrical pads is wire-bonded to a co-designed printed circuit board (PCB) for electrical fan-out [35]. Two lid-less 16-channel 127-µm-pitch polarization-maintaining (PM) fiber arrays are attached to the input and output of the chip using index-matching UV epoxy. Flexible flat cable (FFC) connectors are surface-mounted on the PCB for a compact footprint.

A pluggable SiP MEMS switch array has been demonstrated with contact interposer and ceramic interposer in \cite{hwang2018pluggable}. A $128 \times 128$ MEMS switch with glass interposer and pitch reducing fiber array is demonstrated in \cite{hwang2017128}. A flip chip packaging of a $12 \times 12 $ switch using via less ceramic interposer is reported in \cite{hwang2017flip}.
A $32 \times 32 $ switch with thermo optic MZI and monolithic monitor photodiodes is wirebonded to a CBGA with 68 fiber ribbon in \cite{dumais2017silicon}.

A $32 \times 32$ switch with an extremely high-$\Delta$-PLC connector is demonstrated in \cite{suzuki2019low}.A $32 \times 32$ non duplicate polarization diversity switch is reported in \cite{suzuki2020nonduplicate}. In this paper, authors report switch flip chip bonded to a ceramic interposer that converts 0.18 mm pitch electrodes with 0.5 mm Land Grid Array (LGA), and a 74 port $127 \, \mu m$ fiber array was attached to the chip.

A 3D integrated photonic and electronic chip using $50 \,\mu m$ copper pillar technology is reported in \cite{testa2018integrated}.

An $8 \times 8$ MZI based switch with a submount packaged with 16 narrow core fiber array with SMF on the other side and Spot Size Converter (SSC) on the chip is reported in \cite{nakamura2016optical}. In the same work, authors also report transponder aggregators (TPAs) assembly with 12 co-packaged optical switch modules with current drivers for 1800 thermo optic heaters. 

Hybrid packaging of SiP switches, electronic driver chip and SOA arrays flip chip attached to a silicon carrier is reported in \cite{budd2015semiconductor}. A 12-channel SMF ribbon to chip assembly performed using high-volume tooling with off the shelf components is reported in \cite{lee2018silicon}. A fully integrated $4 \times 4$ switch flip chip integrated with a CMOS driver IC is reported in \cite{rylyakov2011silicon}. A broadband $2 \times 2$ MZI switch is wirebonded to an electronic asic in \cite{lee2013monolithic}. An FPGA embeded system with 16 high speed DAC/ADC and a packaged 4x4 silicon benes switch is reported in \cite{huang2017automated}.A  2 × 2 Mach-Zehnder interferometer (MZI)-based switch element (SE) with active feedback power equalization fabricated in a wafer-scale hybrid silicon process is reported in \cite{chen2013electronic}.

65 nm transistor bulk silicon CMOS technology with a layer of poly-crystalline silicon for photonic devices in a 300 millimeter wafer microelectronics foundry is reported in \cite{atabaki2018integrating}. A zero change process in 32 nm and 45 nm SOI CMOS with both transistors and photonic devices is reported in \cite{stojanovic2018monolithic}. A process in which SOI wafer is oxide bonded face-to-face with the CMOS wafer with Through oxide vias with 3fF parasitic capacitance up to 10x lower than that of micro-bump/copper pillar connection is reported in \cite{stojanovic2015high}.
A 4x4 switch with IBM's 90nm photonics-enabled CMOS process is reported in \cite{dupuis2016cmos}.

\section{Wavelength locking and stabilization}
\label{sec:wavlocking}
Switching elements used in WSS are sensitive to temperature variation and also have to be operated very close to servers in datacenters. Various wavelength locking and stabilization techniques have been reported for MRRs. A feedback control mechanism, based on ContactLess Integrated Photonic Probes and heater actuators, is used to monitor and lock each device in real-time in \cite{zanetto2020wdm}. For multi-wavelength selective crossbar switches, a locking technique which uses heaters in monitor partial power drop ring resonators for calibration is reported in \cite{khope2017chip,khope2017elastic}. In the paper, authors lock two wavelength channels simultaneously from $20-40^{\circ}C$.  Hitless tuning of MRR using a novel channel labelling scheme is reported in \cite{aguiar2018automatic}. A demonstration of locking with photoconductive heaters is reported in \cite{jayatilleka2019photoconductive}. Locking of three rings with a single monitor signal over a temperature range of $>40^{\circ}C$ at 3x20 Gb/s OOK modulation and ~3x75 Gb/s discrete multi-tone (DMT) modulation is reported in \cite{dong2017simultaneous}. An FPGA based tuning algorithm for WDM applications is reported in \cite{gazman2017automated}. A wavelength stabilization scheme for silicon MRR with an in-resonator defect-state-absorption-based photodetector is reported in \cite{li2015active}. A non invasive monitoring scheme using cmos integrated electronics is reported in \cite{grillanda2014non}. A locking scheme based dithering of heater voltage is reported in \cite{padmaraju2013wavelength}. An automated wavelength allignment of a $5^{th}$ order MRR with negligible passband ripples is reported in \cite{mak2015automatic}.
Automatic calibration of 4x4 Benes is reported in silicon\cite{huang2017automated}.

\section{Conclusion}
We summarize the work done on integrated photonic WSS and their applications in datacenters. Electronic and photonic integration has enabled bigger switches with more complicated conntrol circuits. Currently nunmber of ports and number of wavelengths supported are limited by the size of electrical pads. With improvements in packaging technology and fabrication processes, we envision WSS with higher port counts.

\bibliographystyle{unsrtnat}
\bibliography{references}  






\end{document}